\documentclass{emulateapj}

\usepackage{amsmath,natbib, float}

\shorttitle{ML Cluster Mass}
\shortauthors{Ntampaka et al.}
\submitted{Draft version April 29, 2015}

\begin{document}

\title{A Machine Learning Approach for \\ Dynamical Mass Measurements of Galaxy Clusters}

\author{M. Ntampaka\altaffilmark{1}, H. Trac\altaffilmark{1}, D.J. Sutherland\altaffilmark{2}, N. Battaglia\altaffilmark{1, 3}, B. P\'{o}czos\altaffilmark{2}, J. Schneider\altaffilmark{2}}
\email{ntampaka@cmu.edu}
\altaffiltext{1}{McWilliams Center for Cosmology, Department of Physics, Carnegie Mellon University, Pittsburgh, PA 15213}
\altaffiltext{2}{School of Computer Science, Carnegie Mellon University, Pittsburgh, PA 15213}
\altaffiltext{3}{Department of Astrophysical Sciences, Princeton University, Princeton, NJ 08544}
\affil{}

\begin{abstract}
We present a modern machine learning approach for cluster dynamical mass measurements that is a factor of two improvement over using a conventional scaling relation. Different methods are tested against a mock cluster catalog constructed using halos with mass $\geq 10^{14} \ \mathrm{M_\odot}h^{-1}$ from Multidark's publicly-available $N$-body MDPL halo catalog. In the conventional method, we use a standard $M(\sigma_v)$ power law scaling relation to infer cluster mass, $M$, from line-of-sight (LOS) galaxy velocity dispersion, $\sigma_v$. The resulting fractional mass error distribution is broad, with width $\Delta \epsilon \approx 0.8{7}$ (68\% scatter), and has extended high-error tails. The standard scaling relation can be simply enhanced by including higher-order moments of the LOS velocity distribution. Applying the kurtosis as a correction term to $\log(\sigma_v)$ reduces the width of the error distribution to $\Delta \epsilon \approx 0.7{4}$ (1{6}\% improvement). Machine learning can be used to take full advantage of all the information in the velocity distribution. We employ the Support Distribution Machines (SDMs) algorithm that learns from distributions of data to predict single values. SDMs trained and tested on the distribution of LOS velocities yield $\Delta \epsilon \approx 0.4{6}$ ({47}\% improvement). Furthermore, the problematic tails of the mass error distribution are effectively eliminated.  {Decreasing cluster mass errors will improve measurements of the growth of structure and lead to tighter constraints on cosmological parameters.}
\end{abstract}

\keywords{cosmology: theory---dark matter---galaxies: clusters: general---galaxies: kinematics and dynamics---gravitation---large-scale structure of universe---methods: statistical }

\section{Introduction}
\label{sec:intro}

Galaxy clusters have been utilized prominently in astrophysics and cosmology since pioneering work by Fritz Zwicky and George Abell. Clusters are the most massive gravitationally-bound systems in the Universe, with masses $\gtrsim 10^{14}\ \mathrm{M_{\odot}}$, and contain scores to hundreds of galaxies embedded in dark matter halos.  These objects are useful cosmological probes because halo abundance as a function of mass and redshift depend sensitively on fundamental cosmological parameters.  Therefore, measures of cluster abundance can be used to constrain these parameters  {\citep[e.g.][]{2005RvMP...77..207V, 2011ARA&A..49..409A, 2011MNRAS.417.2938S}}.  However, accurately measuring cluster masses for application in cosmology is a difficult endeavor. 

Clusters can be identified across multiple wavelengths.  They were first detected in the visible spectrum as overdensities of galaxies \citep[e.g.][]{1958ApJS....3..211A, 1968cgcg.book.....Z}.  They are identified as overdensities of red galaxies in both visible and IR \citep[e.g.][]{2005ApJS..157....1G, 2010ApJS..191..254H, 2012MNRAS.420.1167A} and can be found as extended X-ray sources \citep[e.g.][]{2002ARA&A..40..539R, 2009ApJ...692.1033V}.  Clusters are also detected by their unique signature in the cosmic microwave background, as a decrement below 218 GHz and an increment above, as predicted by \cite{1972CoASP...4..173S} \citep[e.g.][]{2009ApJ...701...32S, 2011ApJ...737...61M, 2014A&A...571A..29P}.  

Once clusters are identified, mass measurements are needed to map observable cluster properties to the underlying mass.  Cluster masses can be deduced from a variety of techniques including: x-ray observations from which one can infer a mass profile \citep[e.g.][]{2009ApJ...692.1033V, 2010MNRAS.406.1773M}, a temperature-weighted gas mass via the Sunyaev-Zeldovich effect \citep[e.g.][]{2010ApJ...719.1045L, 2013JCAP...07..008H}, mass measurement via strong and weak gravitational lensing 
\citep[e.g.][]{2007arXiv0709.1159J, 2008JCAP...08..006M}, the caustic technique which uses galaxy velocities to determine a mass profile \citep[e.g.][]{2003ApJ...585..205B, 2010MNRAS.407..263A, 2011MNRAS.412..800S}, {the galaxy infall kinematics beyond the virial radius to determine a cluster mass profile \citep[e.g.][]{2013MNRAS.431.3319Z}}
and dynamical mass measurements which employ the virial theorem \citep[e.g.][]{1990ApJS...72..715T, 1996ApJ...458..435C, 1996ApJ...473..670F, 1997ApJ...478..462C, 1998ApJ...505...74G,  2010ApJ...721...90B, 2010ApJ...715L.180R, 2013ApJ...772...25S, 2013arXiv1311.4953R}.   

\cite{zwicky1933rotverschiebung} used the dynamical mass approach.  His work applied the virial theorem, using the dispersion of galaxy velocities to infer the mass of the Coma cluster.  Because dynamical mass measurements probe the entire mass distribution, both regular baryonic matter as well as dark matter, he was able to use the virial theorem to conclude that dark matter outweighed luminous matter in the Coma system.  The virial theorem, which relates kinetic energy to gravitational potential energy, predicts that halo mass relates to galaxy velocity dispersion, $\sigma_v$, as a power law.  

The simplest approach is to treat clusters and their host halos as self-similar, dynamically-relaxed systems with the galaxy velocity dispersion, $\sigma_v$, summing up the halo's dynamics.  However, evidence points toward complications that introduce scatter to the idealized case.
Numerical simulations are useful tools in studying these complex dynamic systems and analyzing sources of scatter in dynamical mass predictions.   
\cite{Evrard:2008aa} find that dark matter particle velocity dispersion scales with total mass as a power law, with a small lognormal scatter.  They identify $\approx10\%$ of the halo population as merger transients, exhibiting higher-than-expected dark matter particle velocity dispersions.  But while a tight correlation is seen between halo mass and the velocity dispersion of simulated dark matter particles $(\sigma_{\mathrm{DM}})$, correlations between the galaxy velocity dispersion ($\sigma_{v}$) and halo mass are more fraught with scatter \citep[e.g.][]{2013ApJ...772...47S}.  A $\lesssim 10\%$ bias between $\sigma_{DM}$ and $\sigma_{v}$ is found, though the sign of this bias is not agreed upon \citep[e.g.][]{Diemand:2004aa, Faltenbacher:2005aa, 2006A&A...456...23B, Faltenbacher:2007aa, 2010ApJ...708.1419L}. 

{Also c}ontributing to the bias and scatter are halo environment and triaxiality {\citep[e.g.][]{White:2010ab, 2013ApJ...772...47S, 2014arXiv1405.0284S, 2013A&A...559A..89W} and projection effects \citep[e.g.][]{2012MNRAS.419.1017C, 2012MNRAS.426.1829N}, while}  \cite{2013MNRAS.434.2606O} find a bias that is dependent on mass and galaxy selection strategy.   {Determining cluster parameters such as center and radii is a difficult endeavor; \cite{2011MNRAS.410..417S} find that the central galaxy may not be the brightest cluster galaxy and additionally is not necessarily at rest with respect to the center of the halo potential well.  Determining which galaxies are cluster members is difficult, and the inclusion of interlopers also introduces scatter \citep[e.g.][]{2010A&A...520A..30M}.}  {\cite{2014MNRAS.441.1513O} compare a variety of cluster mass estimation techniques that rely on galaxy observables, including several virial-theorem-based mass estimates.  They find that scatter in predicted mass increases below halo mass of $10^{14} \, \mathrm{M_\odot}$, as the number of cluster members drops.}  

The complications of triaxiality, environment, galaxy selection, and mergers limit the accuracy with which halo masses can be directly correlated to velocity dispersion.  However, in typical dynamical mass analyses, a wealth of observational information is boiled down in favor of summarizing the halo's dynamics by a summary statistic: $\sigma_v$.  This condensation of information is dictated as much by the virial theorem's $M(\sigma_{v})$ power law relationship as it is by simplicity, because  taking full advantage of the wealth of information available in the full line-of-sight (LOS) velocity probability distribution function (PDF) manually is difficult.  However, the application of nonparametric machine learning (ML) algorithms is a promising resource that may allow cluster masses to be predicted from the myriad data at our disposal.

Machine learning has been applied to astronomy and cosmology problems with mixed results.  Object classification is a common application for machine learning in astronomy, for example: distinguishing Mira variables from semiregular and irregular red variables \citep{2004AJ....128.2965W},  classifying transient events \citep{2008AN....329..288M}, determining galaxy morphology \citep{2010MNRAS.406..342B}{, and more recently, choosing clusters from simulated catalogs which resemble observed clusters \citep{2014arXiv1409.1576H}}.   Other applications range from predicting solar activity \citep[e.g.][]{2009SoPh..255...91Y, 2009SpWea...7.6001C}  to determining photometric redshifts \citep[e.g.][]{2010ApJ...715..823G, 2007ApJ...663..774B}, from cataloging impact craters on Mars \citep{2009Icar..203...77S} to predicting the number of galaxies hosted by a dark matter halo \citep{2013ApJ...772..147X}.

In this work, we focus on improving cluster dynamical mass measurements by moving beyond a simple $M(\sigma_v)$ power law toward more fully utilizing the entire LOS velocity PDF.  In Sec.~\ref{sec:simulation}, we discuss our simulations and mock galaxy catalogs.  In {Secs.~\ref{sec:methods} and \ref{sec:ML}}, we lay out the methods and results for each of three approaches:   first applying a virial-theorem-motivated power law to our mock catalog in Sec.~\ref{sec:PL1}, then using more of the velocity information by taking advantage of higher-order moments in Sec.~\ref{sec:PL2}.  Finally, we utilize the full velocity PDF and the information contained therein by implementing Support Distribution Machines (SDMs), a machine learning algorithm that maps a full LOS velocity distribution to a halo mass prediction in Sec.~\ref{sec:ML}.  We report a comparison of these approaches and a discussion of the interpretation of the results in Sec.~\ref{sec:discussion} and summarize our conclusions in Sec.~\ref{sec:conclusions}.  {We use a $\Lambda$CDM cosmology throughout, with cosmological parameters consistent with Planck data \citep{2014A&A...571A..16P}:  $\Omega_{\Lambda} = 0.69$, $\Omega_m = 0.31$, $\Omega_b = 0.048$, $h = 0.68$, ${n=0.96}$, and $\sigma_8 = 0.82$}

\section{Simulation and Mock Catalog}
\label{sec:simulation}

This work is based on massive halos from the publicly-available Multidark MDPL simulation\footnote{http://www.cosmosim.org/} (Hess et al., in prep.).  
Multidark is an $N$-body simulation containing $3840^3$ particles in a box of length $1\ h^{-1}\rm{Gpc}$ and run using the L-Gadget2 code.  {The mass resolution of this simulation is ${1.51\times10^9\ \mathrm{M_{\odot}} h^{-1}}$, with cosmological parameters consistent with Planck data.}  Massive halos were gleaned from the BDMW halo catalog, which employs a bound density maximum (BDM) spherical overdensity halo finder with halo average density equal to 200 times the critical density of the Universe \citep{1997astro.ph.12217K}.  {These halo masses will be denoted $M$ throughout this work\textbf{.}}   

Halos for our catalog are chosen from the $z=0$ catalog for its large sample of massive halos.  To mimic observed clusters in halo mass, observable minimum luminosity of cluster members (via subhalo mass), and richness, halos are chosen to meet the following criteria:
\begin{enumerate}

 \item Halo minimum mass:  $M \geq 10^{14} \ \mathrm{M_{\odot}} h^{-1}$
 \item Subhalo minimum mass: $M_{\mathrm{subhalo}} \geq 10^{12} \ \mathrm{M_{\odot}}h^{-1}$
 \item Subhalo minimum number: $N_{\mathrm{subhalo}} \geq 20$ 
 
 \end{enumerate}
{The subhalo minimum mass of $M_{\mathrm{subhalo}} \geq 10^{12} \ \mathrm{M_{\odot}}h^{-1}$ corresponds to 663 particles in the smallest subhalos\textbf{.}}
Due to problems with subhalos not being matched with their host halo across the periodic boundary, clusters {with centers that lie} within $2.2 \ \mathrm{Mpc} \,h^{-1}$ of the box edge are pared from the sample.  

{Each subhalo and primary halo in the catalog is assigned a galaxy.  This galaxy is placed at the center of each halo and subhalo, with no galaxy offset considered at this time.  Each galaxy is assigned its subhalo host's velocity.  This method is intentionally simplistic, and is akin to abundance matching with the assumption of zero scatter, i.e.~all subhalos above the minimum subhalo mass host an observable galaxy.  The distribution of galaxies within the cluster is solely determined by the simulated substructure within the primary halo; no assumption about halo mass profiles is made.}  
Galaxy line-of-sight velocities are {calculated} with respect to the mean galaxy LOS velocity of cluster members, with no assumption made concerning which of these galaxies is the brightest cluster galaxy.  Because halo mass is tightly correlated with halo radius, projected subhalo radii ($R_{\mathrm{sub}}$) are normalized by the halo's $R_{200}$.

We prepare {three} catalogs of these clusters.  The {Train C}atalog is used for training and fits, and includes {multiple} line-of-sight views of halos.  {The Train Catalog has a flat mass function by design, and the number of projection per halo in 0.1 dex mass bin varies by bin to achieve 1000 training halos for each bin.}

{The default is to view each cluster first from the cardinal $x-$, $y-$, and $z-$directions and when additional LOS views are needed, these are chosen randomly on the surface of the unit sphere.  While the catalog of unique halos contains 5,028 halos, the Train Catalog contains 15,000 projections of these halos, with more projections of the rare high-mass halos, creating a representative training sample that catalogs how these halos'  velocities and positions might be distributed when viewed from any angle.  }

{Smaller catalogs are used to evaluate the methods, containing} only the three cardinal direction LOS views of each halo.  A mass cut of $M\geq3\times10^{14}\ {M_{\odot}}\, h^{-1}$ is {applied to the Test Catalog} to account for edge effects due to the hard lower mass cut of the sample and selection effects due to the $N_{\mathrm{subhalo}}$ constraint.  {A High-Mass Test Catalog, with a mass cut of ${M\geq7\times10^{14}\ \mathrm{M_{\odot}}\, h^{-1}}$ is also considered.  These catalogs are summarized in Table \ref{table:catalog}}.

\begin{table*}[t]
\begin{center}
\caption{{Catalog Summary} \label{table:catalog}}
\begin{tabular}{l r r l r} 
\tableline
\tableline
\multicolumn{1}{l}{Name} & \multicolumn{1}{l}{Minimum Halo Mass}  & \multicolumn{1}{c}{Unique Halos} & \multicolumn{1}{l}{Projections per Unique Halo} &  \multicolumn{1}{c}{Total Projections} \\
\tableline\\[1ex]
Train Catalog & $1\times 10^{14} \ \mathrm{M_{\odot}}\, h^{-1}$ & 5028 & Mass-dependent \tablenotemark{1} & 15000 \\[1.5ex]
Test Catalog & $3\times 10^{14} \ \mathrm{M_{\odot}}\, h^{-1}$ & 2278 & 3& 6834\\[1.5ex]
High-Mass Test Catalog & $7\times 10^{14} \ \mathrm{M_{\odot}}\, h^{-1}$ & 315 & 3 & 945 \\[1.5ex]

\tableline
\end{tabular}

\tablenotetext{1}{To create a flat mass function with an equal number of training points per 0.1 dex halo mass bin. }

\end{center}
\end{table*}

It should also be noted that, despite the {multiple projections of high-mass halos in the Train Catalog,} there can be challenges at the high-mass end of the training sample. Training and predicting on the halos toward the more massive end of the spectrum is difficult because they are rare, leading to large statistical uncertainties.

\section{{Power Law Methods and Results}}
\label{sec:methods}

In {the next two sections,} we move through successive methods, first describing the method, next reporting the results, then adding a layer of complexity with the subsequent method.  Motivated by the virial theorem, we will explore improvements to dynamical mass measurements.  The methods will be described in detail in the following {sections}, but are summarized in Table \ref{table:methods} for reference.  They include the virial-theorem-motivated $M(\sigma_v)$ power law (PL1), a power law method that includes applying higher-order moments of the LOS velocity data (PL2), and four machine learning methods, each employing the same ML algorithm with different inputs:  LOS velocities only (ML1 and ML2), and LOS velocities plus galaxy projected sky position (ML3 and ML4).

\begin{table*}[t]
\begin{center}
\caption{Methods \label{table:methods}}
\begin{tabular}{l l l l l} 
\\
\tableline
\tableline
Case & Approach & Summary Stats & Distribution Features & Color \\ [0.5ex] 
\tableline\\[1ex]
PL1 & Power Law & $\sigma_v$ &---& Red \\ 
PL2 & Power Law  &$\sigma_v$ \& $\kappa$ &---& Blue \\ 
ML1 & Machine Learning: SDM  &---& $|v_{\mathrm{los}}|$  & Green \\ 
ML2 &  Machine Learning: SDM &---& $|v_{\mathrm{los}}|$ \& $|v_{\mathrm{los}}|/\sigma_v$ & Purple \\ 
ML3 &  Machine Learning: SDM  &---& $|v_{\mathrm{los}}|$ \& $R_{\mathrm{sub}}/R_{\mathrm{200}}$ & Orange \\ 
ML4 & Machine Learning: SDM   &---& $|L_{\mathrm{eff}}|$ & Brown \\ [2ex] 
\tableline
\end{tabular}
\end{center}
\end{table*}

\subsection{$M(\sigma_v)$ Power Law}
\label{sec:PL1}

We begin by using the virial theorem as a jumping-off point for dynamical mass measurements.  The virial theorem states that, for an object that is stably bound by gravity, $2K+U=0$, where $K$ is the kinetic energy of the system and $U$ is its gravitational potential energy.  This can be extended to find a relationship between velocity dispersion, $\sigma_v = \left<v^2\right>^{1/2}$, and mass, $M$:
\begin{equation}
\label{eq:sigma}
\sigma_v^2 = \alpha_R \frac{ G M}{R},
\end{equation}
where $\alpha_R$ is a constant of order one that depends on the cutoff definition of halo radius R and the density profile of the dark matter halo.  It can be constrained by observation \citep[e.g.][]{1969ApJ...158L.139S}.  For a halo population with a known average mass density, $M\propto R^{3}$, and velocity dispersion relates directly to mass as
\begin{equation}
\label{eq:sigM}
\sigma_v \propto M^{1/3}.
\end{equation}
Studies of $N$-body simulations find that this power law generally holds with slope $\approx 0.33$ \citep[e.g.][]{Evrard:2008aa}.  Although there is a tight $M(\sigma_{DM})$ relationship for the velocity dispersion of dark matter, the halo mass is less tightly correlated with $\sigma_v$, the velocity dispersion of galaxies \citep[e.g.][]{2013ApJ...772...47S}.  

Halo mass $M$ can be related to galaxy LOS velocity dispersion $\sigma_v$ via the power law
\begin{equation}
\sigma_{v}(M) = \sigma_{15}\left( \frac{ M}{10^{15} \, \mathrm{M_\odot}  h^{-1}} \right)^\alpha.
\label{eq:powerlaw}
\end{equation}
We find least-squares fit to $\log(\sigma_v)=\alpha\log(M)+\beta$ for the Train Catalog, binned in 0.1 dex mass bins.  The power law best fit parameters {for the Train Catalog} are $\alpha=0.3{82}$ and $\sigma_{15} = 12{44} \ \mathrm{km\, s^{-1}}${; this result is relatively insensitive to the catalog choice, and comparable best fit values are calculated for the Test Catalog.  However, w}e caution that these parameters are a fit for a particular simulation and catalog, and should be applied with care to predict cluster masses for observational data.  

\begin{figure}[tb]
  \centering
  \includegraphics[width=0.5\textwidth]{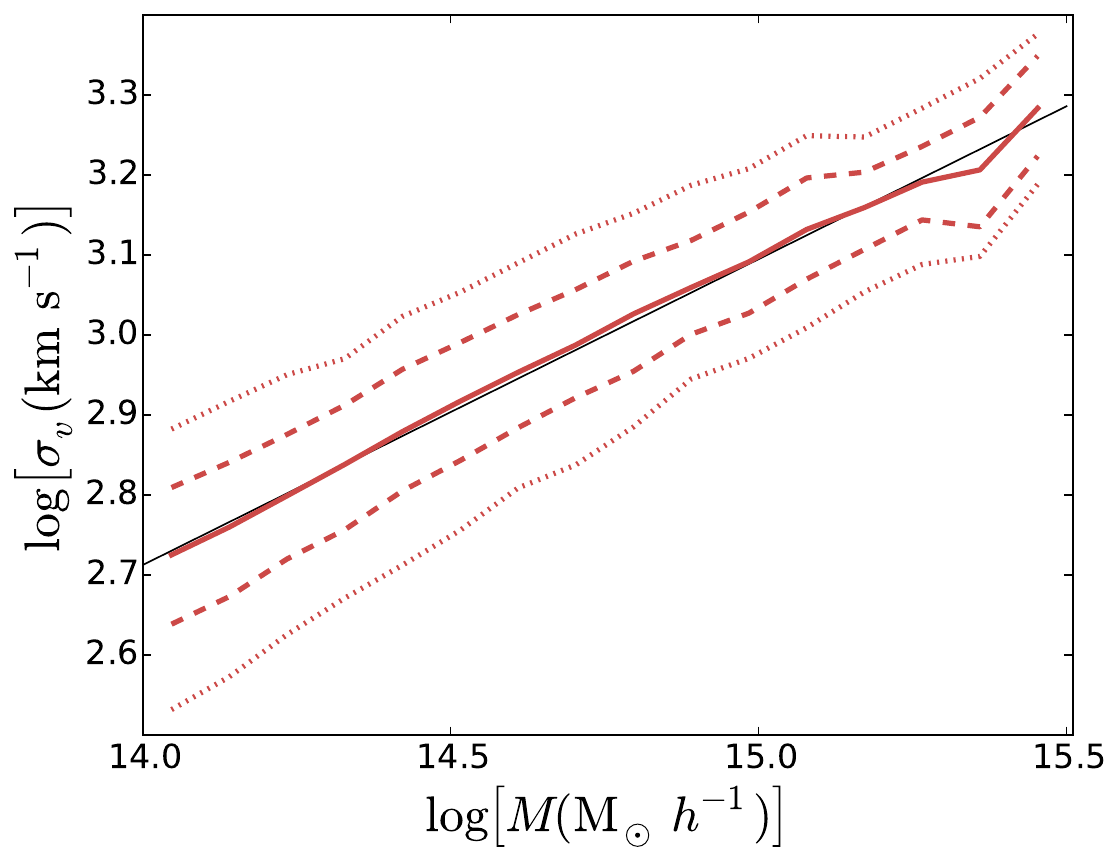}
  \caption{Line-of-sight galaxy velocity dispersion, $\sigma_v$, {vs}.~halo mass, $M$.  The binned halos are fit by a power law relation (thin solid black line).  The 68\% (dashed) and 95\% (dotted) scatter show the substantial uncertainty in this $M(\sigma_v)$ relationship.}

  \label{fig:Msig}
\end{figure}

Figure \ref{fig:Msig} shows the power law best fit and both 68\% and 95\% scatter.  The power law is a good fit for median binned $\sigma_v$, albeit with significant scatter at all mass ranges considered.

It should be noted that the best fit to an $M(\sigma_{DM})$ relationship for this simulation has $\alpha = 0.34$, much closer to the expected value of 0.33.  The steepness of our best fit slope can be attributed to both numerical and physical effects.  \cite{2013ApJ...772...47S} find that measurements of $\sigma_v$ from small samples (e.g.~$\lesssim 40$ galaxies) of very luminous galaxies tend to be smaller than the $\sigma_v$ calculated with the inclusion of more plentiful, fainter galaxies.  Because the halos at the low mass end of our catalog tend to have the fewest galaxies (on the order of $20$ per cluster), this trend would preferentially bias the velocity dispersions down for the smallest-mass clusters.  The lower-than-expected dispersions at the low-mass end thus steepen the best fit power law, giving an $\alpha$ greater than the expected value of $0.33$.  Additionally, \cite{2013MNRAS.430.2638M} find that dynamical friction and tidal disruption affect subhalo $\sigma_v$, with tidal stripping having a greater effect on higher-mass clusters, creating a velocity bias that changes with cluster mass.  Both numerical and physical effects may come into play in our catalog, where a steeper $\alpha$ is evident for low-mass systems and a shallower one for higher-mass clusters.  

The distribution of $\sigma_v$ is roughly log-normal for halos of a given mass, and we define the residual, $\delta$, as
\begin{equation}
\label{eq:delta}
\delta =\log( \sigma_{v})-\log(\sigma_{v, \, \mathrm{best\,fit}}),
\end{equation}
where $ \sigma_{v}$ is the velocity dispersion calculated from LOS velocities and $\sigma_{v, \,\mathrm{best\,fit}}$ is the expected velocity dispersion for a given halo mass (i.e.~the velocity dispersion that would be calculated by Equation \ref{eq:powerlaw} with the $\alpha$ and $\sigma_{15}$ values given).  

The biweight estimator given by \cite{1990AJ....100...32B} was considered for this and subsequent methods.  However, {a strength of the biweight estimator is in detecting and omitting interlopers' contribution to the sample variance; these interlopers have already been excluded by the catalog design.  This estimator} was found to have a larger range $\delta$ values compared to when the velocity dispersion was defined as the standard deviation.  
{In order to provide a comparison to the most idealized power law method}, the standard deviation was chosen as a measure of velocity dispersion, $\sigma_v$, for this catalog.

\subsection{$M(\sigma_v)$ Power Law with Kurtosis}
\label{sec:PL2}

In the simple $M(\sigma_v)$ power law applied in Sec.~\ref{sec:PL1}, the full information of the LOS velocity PDF is summarized by a single statistic.  But this PDF contains more information than is used.  In the upcoming section, we will explore using higher-order moments of the PDF to improve dynamical mass measurements of galaxy clusters.  

\begin{figure}[t]
  \centering
  \includegraphics[width=0.5\textwidth]{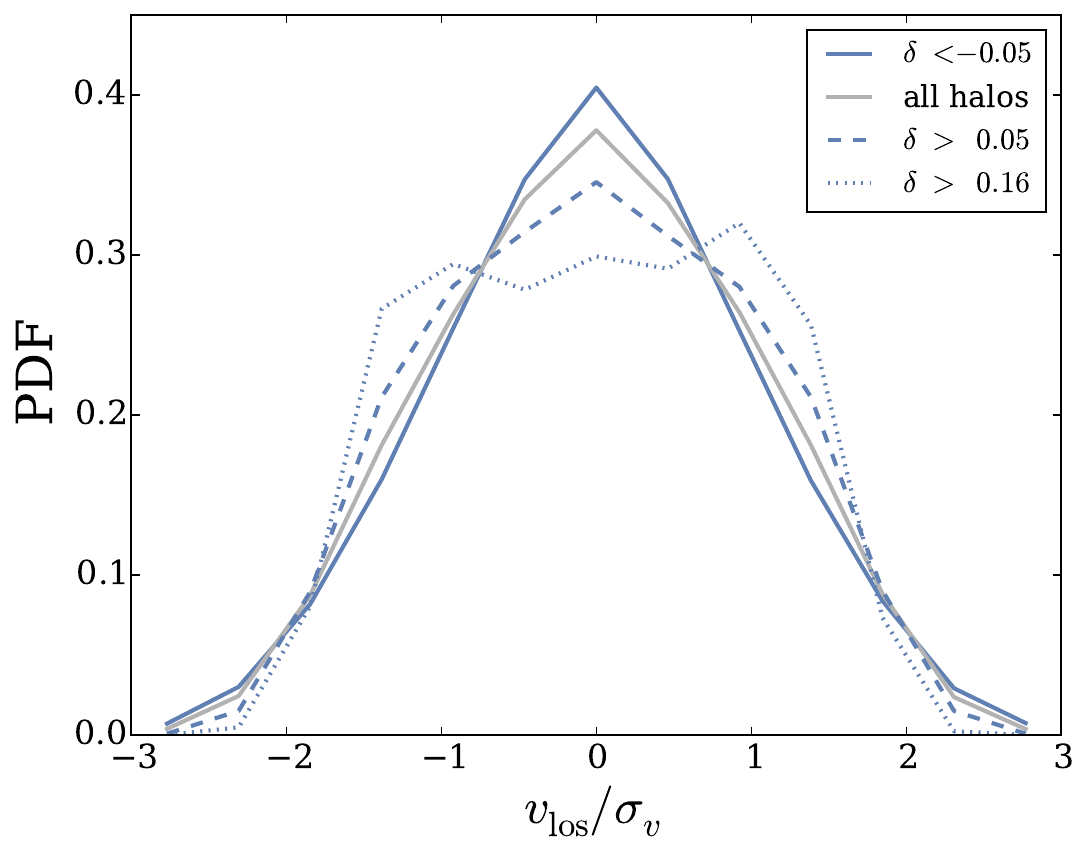}
  \caption{PDFs of normalized galaxy LOS velocities from stacked halos.  The {Test Catalog} halo population (gray solid) is roughly Gaussian.  However, halos with a large negative $\delta$ and underpredicted masses exhibit a sharply peaked PDF (blue solid), while those with a large positive $\delta$ and overpredicted masses have a flatter PDF (blue dashed).  In the most extreme large positive $\delta$ (blue dotted), a flat, wide PDF emerges.  The shape of the velocity PDF, quantified by the kurtosis, can be used to predict $\delta$.
}

  \label{fig:stacked}
\end{figure}

Figure \ref{fig:stacked} shows that stacked halos from differing populations---the full Test Catalog, large positive $\delta$, and large negative $\delta$---exhibit strikingly different shapes.  While the full Test Catalog's shape is approximately Gaussian, the halos with a large negative $\delta$ are more sharply peaked near $v_{\mathrm{los}}/ \sigma_v=0$ and the halos with large positive $\delta$ have a flatter distribution.  

This difference in shape---the sharply peaked compared to the overly-flat curves---can be quantified by the excess kurtosis, $\kappa$, defined as:
\begin{equation}
\kappa=\frac{\sum\limits_{i=1}^N (v_i-\overline{v})^4}{N \sigma_{v}^4}-3,
\end{equation}
where $v_i$ is the line-of-sight velocity of the $i^{th}$ galaxy, $\overline{v}$ is the mean line-of-sight velocity of galaxies in the cluster, $N$ is the number of galaxies in the cluster, and $\sigma_{v}$ is the standard deviation  of a cluster's galaxy line-of-sight velocities.  The subtraction of 3 sets this value such that a standard normal distribution has an excess kurtosis of 0.  Throughout this work, we refer to excess kurtosis simply as ``kurtosis.''

\begin{figure}[h]
  \centering
  \includegraphics[width=0.5\textwidth]{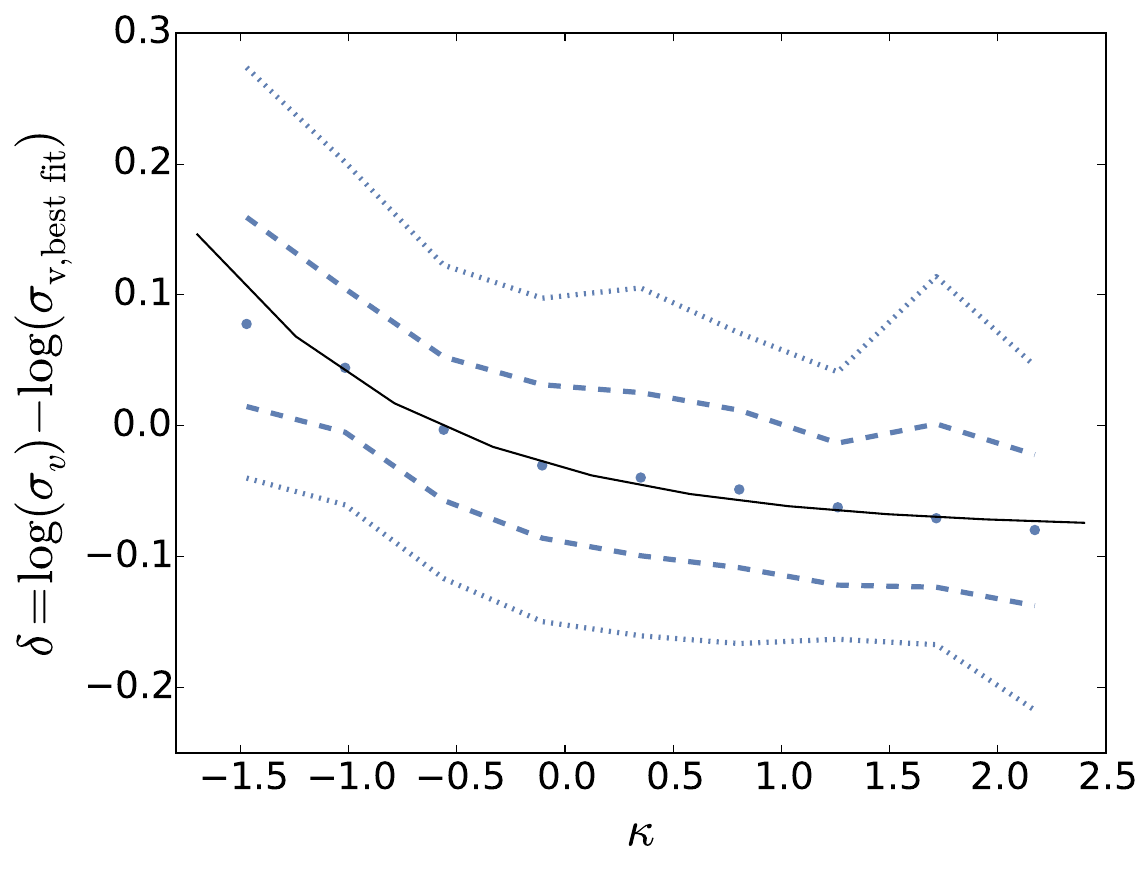}
  \caption{The kurtosis, $\kappa$, of the LOS velocity PDF is a predictor of $\delta$ (the residual from the $M(\sigma_v)$ power law, given by Equation \ref{eq:delta}).  Individual clusters are binned, and the binned mean $\delta$ (blue points) are fitted to a decaying exponential (solid black curve).  {Dashed} lines show binned 68\% scatter; {dotted} show 95\% scatter.  Exploiting this relationship allows us to predict $\delta$ based on the kurtosis of the LOS velocity PDF and produce a $\kappa$-corrected $\sigma_v$ to improve halo mass predictions.}

  \label{fig:kurtosis}
\end{figure}

Expanding on the illustration in Figure \ref{fig:stacked}, Figure \ref{fig:kurtosis} shows the relation between $\delta$ and $\kappa$.  Halos with wide, flat velocity PDFs (i.e.~those with negative $\kappa$) tend to have larger $\delta$ values, while halos with sharply-peaked, high-$\kappa$ PDF's have moderate-to-low $\delta$ values.  The simplest reasonable fit for median $\delta$ as a function of binned mean kurtosis is a decaying exponential,
\begin{equation}
\delta(\kappa)= a\exp(-b*\kappa)+c,
\label{eq:deltakappa}
\end{equation}
with best fit parameters $a=0.04{6}$, $b=0.9{3}$, and $c=-0.07{9}$. Using this fit as a predictor for residual $\delta$ in equation \ref{eq:delta} allows us to calculate a $\kappa$-corrected velocity dispersion ($\sigma_{v,\,\mathrm{\kappa\mbox{-}corrected}}$), 

\begin{equation}
	\log(\sigma_{v,\, \mathrm{\kappa\mbox{-}corrected}}) = \log(\sigma_{v})-\delta(\kappa),
	\label{eq:sigkap}
\end{equation}
from the measured velocity dispersion, $\sigma_v$, and the residual, $\delta(\kappa)$.  

We now find a power law relating $M$ and $\sigma_{v,\, \mathrm{\kappa\mbox{-}corrected}}$ for the Train Catalog.
Correcting the $\log(\sigma_v)$ values linearly with $\delta(\kappa)$ (as in equation \ref{eq:sigkap}) does not significantly alter the power law fit, with $\kappa$-corrected best fit values to Equation \ref{eq:powerlaw} of $\alpha=0.38{4}$ and $\sigma_{15} = 12{41} \  \mathrm{km\, s^{-1}}$.  

We next consider the fractional mass error
\begin{equation}
	\epsilon=(M_{\textrm{pred}}-M)/M,
	\label{eq:fracerror}
\end{equation}
comparing the actual halo mass, $M$, to the predicted halo mass, $M_{\textrm{pred}}$.  When applying the $\kappa$-corrected power law to halos in the Test Catalog, the fractional error decreases.  The mean $\epsilon$ has moved closer to zero, from ${0.128}$ to $0.07{9}$.   The width of the 68\% scatter in fractional mass error decreased as well, from $0.8{7}$ to $0.7{3}$.  The inclusion of $\kappa$ as a correction term has allowed us to better predict halo masses from our line-of-sight velocities, reducing both bias and scatter in fractional mass error.

\begin{figure}[!tb]
  \centering
  \includegraphics[width=0.45\textwidth]{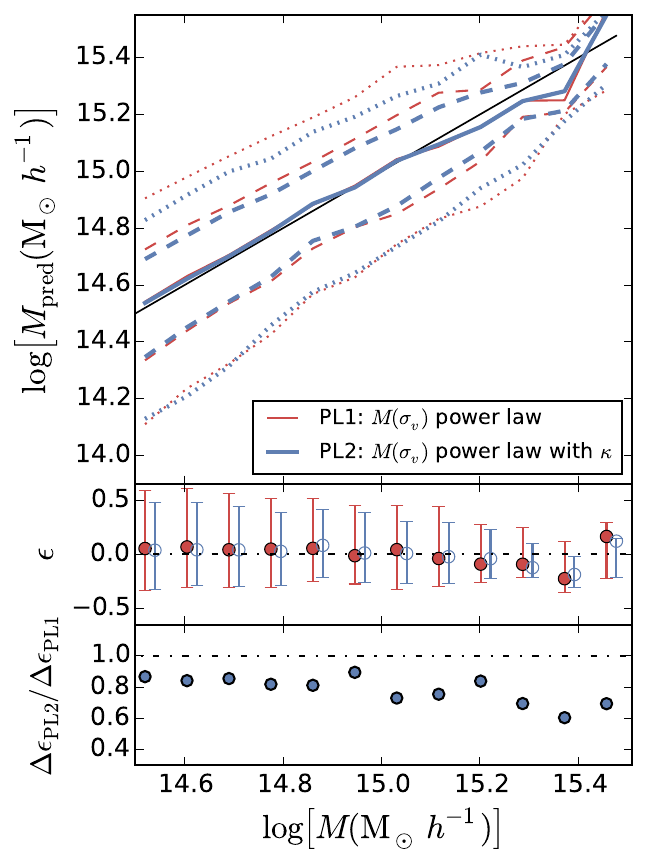}
  \caption{ By adding kurtosis as a predictor of residual, we are able to decrease the scatter in the $M(\sigma_v)$ relationship.
Top panel:   predicted halo mass, $M_{\mathrm{pred}}$ vs.~halo mass, $M$.  
PL2 mass predictions are improved by the addition of kurtosis to the $M(\sigma_v)$ power law.  The PL2 method (thick blue) shows an improvement over the PL1 method (thin red).  Binned median (solid), 68\% (dashed) and 95\% (dotted) scatter are shown for both PL1 and PL2.
Middle panel:  fractional mass error for PL1 (filled red circles) and PL2 (open blue circles).  Points are the median $\epsilon$, and error bars show 68\% scatter.  Blue points are displaced horizontally for clarity.  The addition of kurtosis as a linear correction term slightly improves the negative bias at high masses, and decreases scatter at all mass scales.
Bottom panel:  the width of the 68\% scatter of PL2 ($\Delta\epsilon_{\mathrm{PL2}}$) is compared to that of PL1 ($\Delta\epsilon_{\mathrm{PL1}}$).  Kurtosis as a predictor of $\delta$ decreases the width of $\epsilon$ for all mass bins. }

   \label{fig:PLplusK}
\end{figure}

Figure \ref{fig:PLplusK} shows the overall improvement of halo mass measurement of  $\sigma_{v,\ \kappa\mathrm{\mbox{-}corrected}}$ as compared to the measured velocity dispersion, $\sigma_v$.  The top panel shows that at all mass ranges, the 68\% and 95\% scatters have decreased.  The middle panel shows that the median fractional mass error, $\epsilon$, edges closer to a zero bias  for most bins $\geq10^{15}\ \mathrm{M_\odot}h^{-1}$.  {Halos with mass $\geq10^{15}\ \mathrm{M_\odot}h^{-1}$ are found to have a more tightly correlated $\kappa-\delta$ relationship, which may explain why this $\kappa$ correction method preferentially improves the prediction of the highest-mass halos.}  The bottom panel quantifies the ratio of 68\% scatter of PL2 ($\Delta\epsilon_{\mathrm{PL2}}$) compared to the 68\% scatter of PL1 ($\Delta\epsilon_{\mathrm{PL1}}$), showing that with the inclusion of $\kappa$ in our $M(\sigma_v)$ power law, the scatter in $\epsilon$ decreases.

The application of kurtosis, $\kappa$, as a predictor for $\delta$ results in an improvement of mass measurements across masses considered.  This result should not be surprising:  kurtosis profiles have been used in conjunction with the Jeans equation to explore mass profiles \citep[e.g.][]{2003MNRAS.343..401L, 2006MNRAS.367.1463L}; we refer to the original papers for details on this approach.  

Mergers and infalling matter offer a natural explanation for the correlation between high $\delta$ and negative $\kappa$. A halo undergoing a merger or experiencing infalling matter tends to have a flat-topped or double-peaked velocity distribution \citep[e.g.][]{2011MNRAS.413L..81R}, resulting in a negative velocity PDF kurtosis.  This corresponds well with what is found from simulated dark matter particle dispersion, that halos undergoing a merger tend to have masses overpredicted by an $M(\sigma_{\mathrm{DM}})$ relationship \citep{Evrard:2008aa}, leading to a large positive residual, $\delta$.  In the case of negative kurtosis, the relationship between $\kappa$ and $\delta$ seems to be driven by the identification of halos undergoing a period of mass growth. 

It should be noted that other moments of the LOS velocity PDF may be applied in a similar manner.   Skewness, for example, is also a weak predictor of $\delta$ and can also be used as a correction term to the $M(\sigma_v)$ power law.  For our catalog, applying skewness as a correction term to the $M(\sigma_v)$ power law decreases the mean fractional mass error by $\approx{22}\%${, reducing the tendency to overpredict halo mass that is seen in PL1.  However, the application of skewness as a correction term makes} no significant {decrease} in the 68\% scatter.  

Though we have improved dynamical mass measurement by taking more advantage of the information encoded in the LOS velocity PDF, the method we have used here still merely summarizes the full velocity PDF: we have moved from one summary statistic, $\sigma_v$, in PL1, to two, $\sigma_v$ and $\kappa$, in PL2.

\section{{Machine Learning Methods and Results}}
\label{sec:ML}

\begin{figure*}[t]
  \centering
  \includegraphics[width=\textwidth]{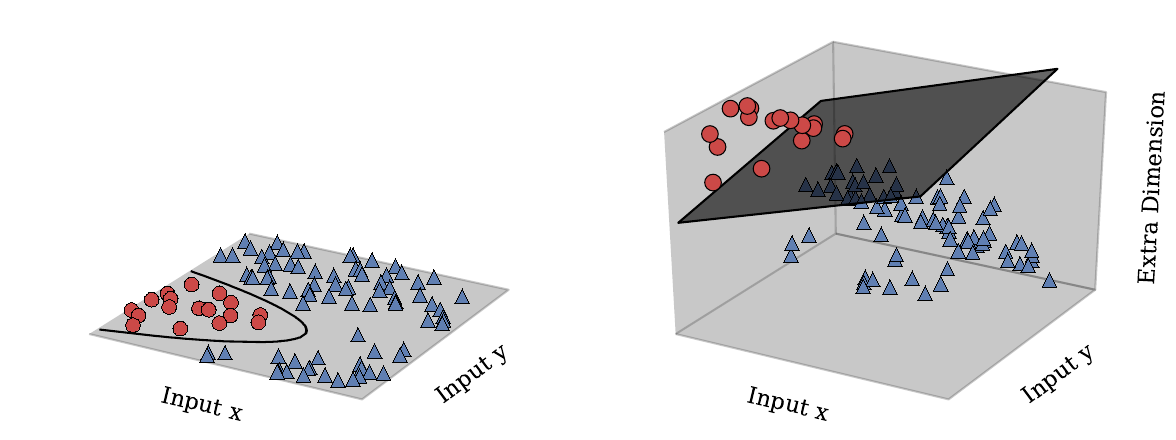}
  \caption{In their basic form, Support Vector Machine classifiers search for a hyperplane that separates two classes, here shown as red circles and blue triangles. In the two-feature input space (left panel), the two classes are not separable by a hyperplane. But if the inputs are re-cast into an appropriate higher-dimensional space (right panel), the red circles and blue triangles can be divided by a plane. Test examples can now be classified based on the side of the plane on which they lie; a test example lying above the plane would be classified as a red circle, and one below as a blue triangle. The ``decision boundary'' which divides the predicted classes (black plane, right panel) corresponds to a different shape in the original space (solid black curve, left panel). We can find the boundary in the original space without explicitly mapping to the higher-dimensional space using the ``kernel trick.'' Although we wish to predict a continuous-valued label rather than a binary classification, an analogous process applies.}
  \label{fig:kernelmachine}
\end{figure*}

Moving beyond summary statistics toward utilizing the full information encapsulated in the velocity PDF cannot easily be done manually.  Taking advantage of the full LOS velocity PDF is a good candidate for moving from simple power law relationships to machine learning.  The problem at hand is to take samples from a probability distribution (galaxy line-of-sight velocities and projected positions) and map these to a final numerical prediction (halo mass).  Support Distribution Machines\footnote{https://github.com/djsutherland/py-sdm} (SDMs)  \citep{2012arXiv1202.0302S} are chosen as candidates to solve this problem for their virtue of learning from a distribution and predicting a scalar.  {In this section, we present the Support Distribution Machines framework, implementation, and the results of this machine learning approach applied to halo mass prediction.}

\subsection{{SDM Theory}}

SDMs are built upon Support Vector Machines (SVMs).  SVM is a machine learning method that, in its simplest form, takes a set of training example data, with data vectors from which to learn, and divides them such that training data with similar labels lie on the same side of the boundary.  Figure \ref{fig:kernelmachine} gives a simple illustration of how training examples with two features might be separated by a decision boundary.  Test examples are then classified according to which side of the decision boundary they fall on.  The algorithm we use here differs from an SVM classifier in two key ways.  First, rather than simply classifying test examples into a finite number of categories, the output values are real numbers; this is called support vector regression (SVR).  Second, SDMs learn from distributions rather than from data vectors.

The formal description of SVR is as follows \citep{Drucker97supportvector,scholkopf2002learning}.  In this supervised learning problem we have $\{(X_n,Y_n)\}_{n=1}^N$ (input, output) pairs, where  $Y_n\in \mathbb{R}$, $X_i\in\mathbb{R}^d$.  The primal form of support vector regression is formulated as a convex optimization problem:
\begin{align}
&\min \|w\|^2 \label{eq:SVR-primal-hard}\\
&\textrm{subject to} \begin{cases}Y_n-w^T\phi(X_n) <\epsilon_{\mathrm{tol}} \\w^T\phi(X_n)-Y_n<\epsilon_{\mathrm{tol}} \end{cases},\nonumber
\end{align}
where $w\in\mathbb{R}^D$, $\phi: \mathbb{R}^d \to \mathbb{R}^D$ is a user-defined feature map, and $\epsilon_{\mathrm{tol}}>0$ is a user-defined error tolerance parameter. 
The intuition behind these equations is that we want to find a linear map with small weights ($w$) such that, in the training points, the regression error is smaller than the parameter $\epsilon_{\mathrm{tol}}>0$.

Depending on the data and parameters, these constraints can easily be infeasible, therefore, analogously to the ``soft margin'' loss function \cite{bennett1992robust}, which was adapted to SVM by \cite{cortes1995support}, one can introduce slack variables $\xi_n, \xi_n^*$ to relax the equation \eqref{eq:SVR-primal-hard}. After introducing these slack variables, we
arrive at the following primal convex problem stated by \cite{vapnik2000nature}:

\begin{align} \label{eq:SVR-primal}
&\min \bigg( \|w\|^2+C \sum_{n=1}^N (\xi_n+ \xi_n^*) \bigg) \\
&\textrm{subject to} \begin{cases}Y_n-w^T\phi(X_n) <\epsilon_{\mathrm{tol}}+\xi_n \\w^T\phi(X_n)-Y_n<\epsilon_{\mathrm{tol}}+\xi_n^*\\ \xi_n,\xi^*_n \geq 0,\end{cases} \nonumber
\end{align}
where $C>0$ is a parameter. Instead of directly solving the primal quadratic problem \eqref{eq:SVR-primal}, in many applications it is easier to solve its dual problem instead:
\begin{align} \label{eq:SVR-dual}
&\max \bigg(  -\frac{1}{2} \sum_{i,j}^N (\alpha_i-\alpha_i^*)(\alpha_j-\alpha_j^*)k(X_i,X_j) \\
&-\epsilon_{\mathrm{tol}}\sum_{n=1}^N (\alpha_n+\alpha_n^*)+\sum_{n=1}^N Y_n(\alpha_n-\alpha_n^*) \bigg) \nonumber \\
&\textrm{subject to} \sum_{n=1}^N (\alpha_n-\alpha_n^*)=0, \textrm{and } 0\leq \alpha_n, \alpha_n^* \leq C. \nonumber
\end{align}
Here, $k(X_i,X_j)\doteq \phi(X_i)^T\phi(X_j)$ is a so-called ``kernel function.'' The predicted value for a new input $X$ is given by
$f(X)=\sum_{n=1}^N(\alpha_n-\alpha_n^*)k(X_n,X)$. A frequently used kernel function is the Gaussian kernel $k(a,b)=\exp(-\|a-b\|^2/\sigma^2)$, for some $\sigma>0$ parameter, but any positive semi-definite (PSD) function can be used as kernel $k(\cdot,\cdot)$.

One crucial difference between SVR and our problem is that, in our case, the input $X_n$ is not a finite-dimensional vector, but a distribution with density function $p_n$. For the kernel value between distributions $p_i$ and $p_j$, we will use $k(p_i,p_j)\doteq \exp(-KL(p_i,p_j)/\sigma^2)$, where $KL(p_i,p_j)=\int p_i(x)\log(p_i(x)/p_j(x))$ is the Kullback\textendash Leibler (KL) divergence.\footnote{To make this kernel $k$ PSD, we project the Gramm matix $G$ of the data ($G_{ij}=k(X_i,X_j)$) to the closest PSD matrix in Frobenius norm.} In our problem, of course, we do not know these densities exactly; only sample sets are available to us. We will use these sample sets to estimate the  KL divergence using the estimator given by \cite{4839047}. The KL divergence estimate, $\textrm{KL}_{n, m}(X_A || X_B)$, from the feature-space samples $X_A$ from Cluster $A$ (containing $n$ galaxies) to the $X_B$ samples from Cluster $B$ (containing $m$ galaxies) is given by
\begin{equation}
\label{eq:KL}
	\textrm{KL}_{n, m}(X_A || X_B) = \frac{d}{n} \sum_{i=1}^{n}\log{\frac{\nu_k(i)}{\rho_k(i)}}+\log\frac{m}{n-1}
\end{equation}
where $d$ is the number of dimensions, i.e.~the number of distribution features considered in the method, $\nu_k(i)$ is the Euclidean distance in input space from the $i^{th}$ galaxy in $X_A$ to its k-nearest neighbor in $X_B$ and $\rho_k(i)$ is the distance from the $i^{th}$ galaxy in $X_A$ to its k-nearest neighbor in $X_A$.  We use $k=2$ throughout. 

The training catalog is first used to select kernel parameters $C$ and $\sigma$ via 3-fold cross-validation.  It is then used to train the regression model with the best-selected kernel, which in turn is used to predict the masses of the clusters in the Test Catalog.  For further information on the Support Distribution Machine regressor, see \cite{2012arXiv1202.0302S}.

\subsection{{SDM Implementation}}

\begin{figure*}[!htb]
\begin{center}
\begin{tabular}{c c}
     
     ML1 & ML2\\
     \includegraphics[width=0.38\textwidth]{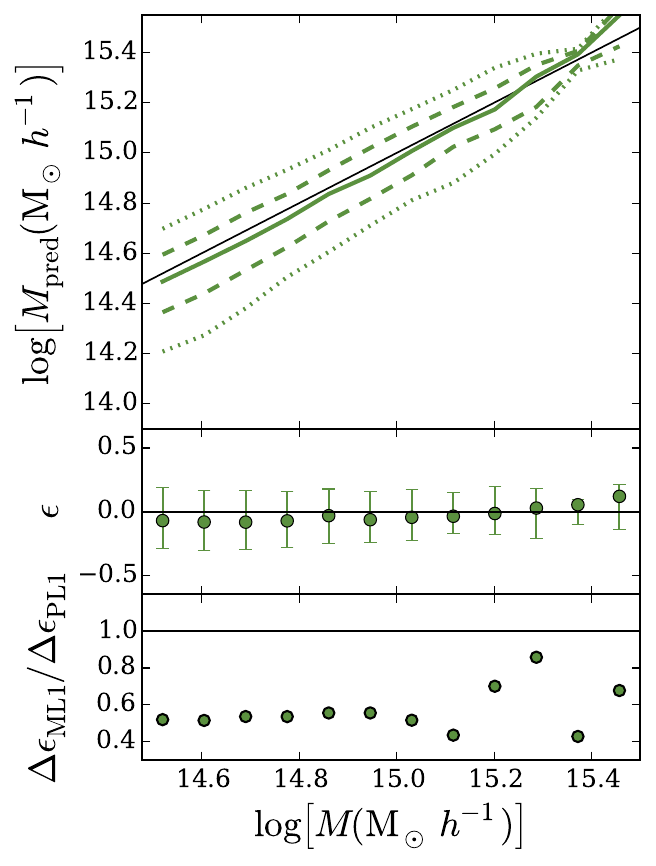} & \includegraphics[width=0.38\textwidth]{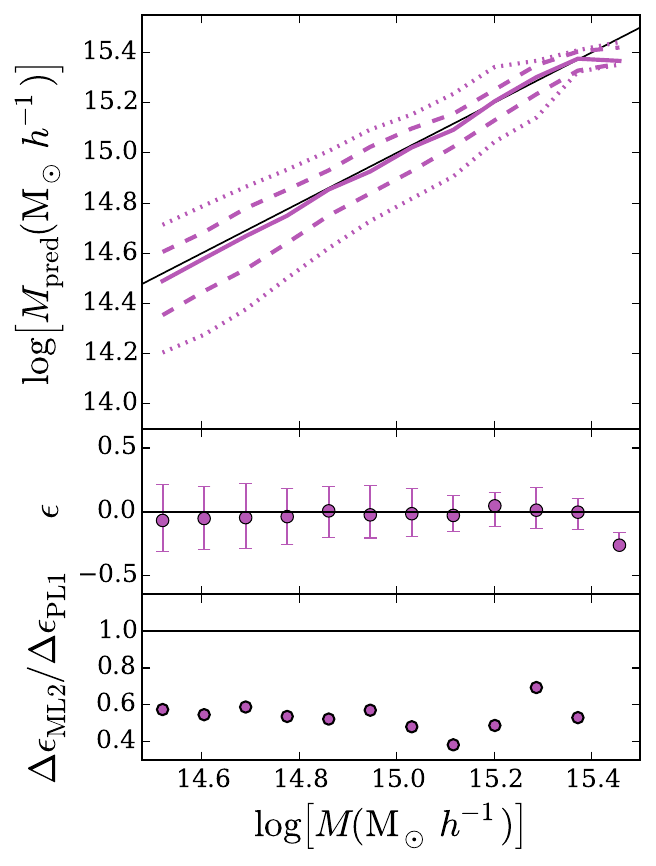} \\ [1.5mm]
    ML3 & ML4 \\
      \includegraphics[width=0.38\textwidth]{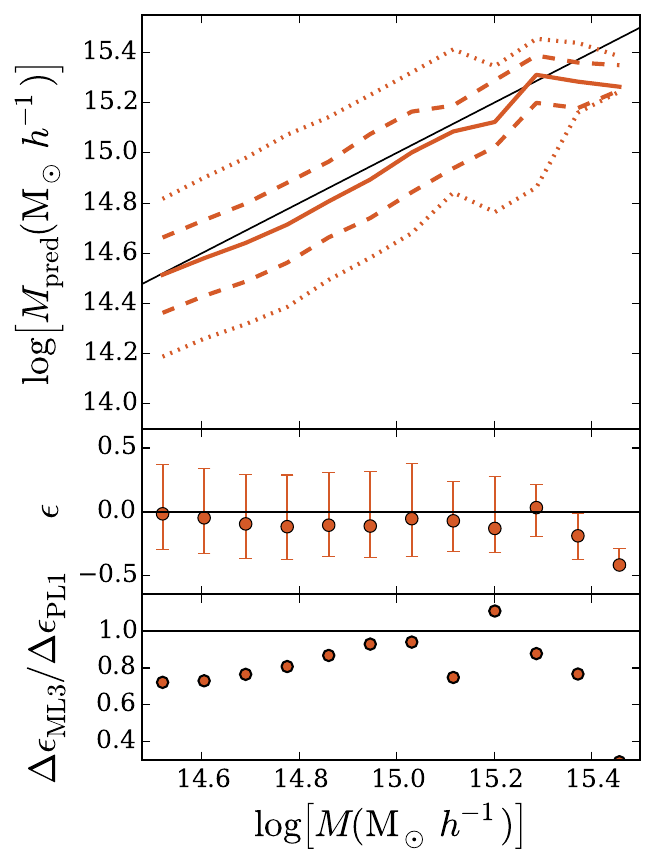} &  \includegraphics[width=0.38\textwidth]{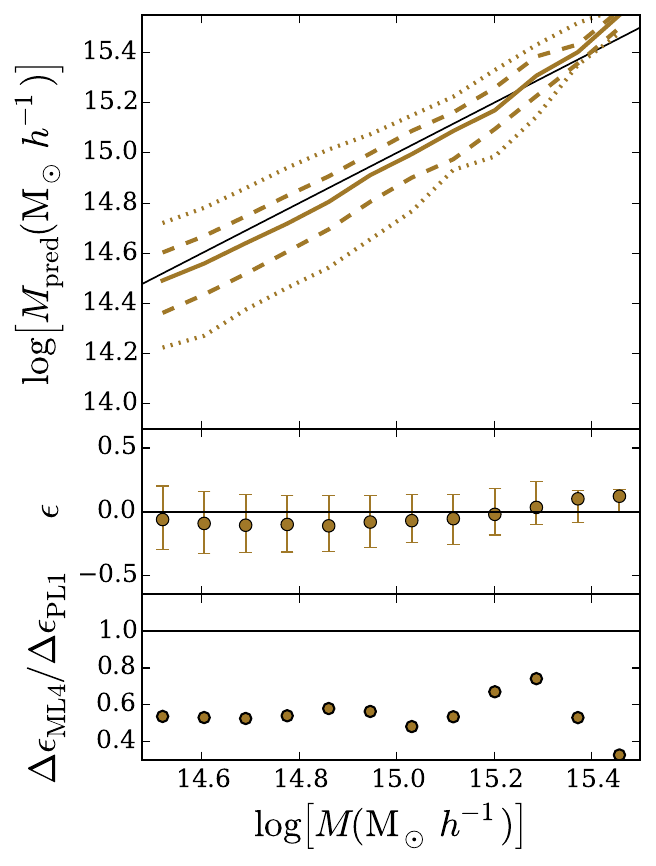}
      
  \end{tabular}
      \caption{\footnotesize{Summary of the SDM results.  Four large panels show the four machine learning approaches:  ML1 (top left, green), ML2 (top right, purple), ML3 (bottom left, orange), ML4 (bottom right, brown).    
            Within each large panel, three sub panels show: 
      Top panel:   Halo predicted mass, $M_{\mathrm{pred}}$, vs.~actual mass, $M$.  The binned predicted mass median, 68\%, and 95\% scatter (solid, dashed, and dotted, respectively) are also shown.
  Middle panel:  Fractional mass error as a function of halo mass.  Points are the median $\epsilon$ and error bars show 68\% scatter.    
  Bottom panel:  Error width, $\Delta\epsilon_{\mathrm{ML}}$ relative to that of PL1 (virial-theorem-motivated power law).
  Three methods---ML1, ML2, and ML4---show pronounced improvements over the $M(\sigma_v)$ power law (PL1).}}
      \label{fig:MLsummary}
      \end{center}
\end{figure*}

Before applying the SDM regressor, the {unique halos are} rank-ordered by mass and alternately divided into {ten} data sets{, or folds.  The Train Catalog data for nine of these folds is used to train the SDM fit, and the resulting fit is used to predict the Test Catalog halos in the tenth, unused fold.  This process is repeated ten times, using multiple projections of 90\% of the unique halos to train, and predicting on three LOS views of the remaining 10\% of unique halos.  In this way, each of the three views of each unique halo in the Test Catalog is predicted.}

To explore how both the line-of-sight galaxy velocities as well as their relative positions might affect mass predictions, four sets of training features are considered.  Each model uses one or two distribution features, implementing constructs of the sets of $v_{\mathrm{los}}$ and/or $R_{\mathrm{sub}}/R_{200}$ only.  See Table \ref{table:methods} for a summary of the features used in each method.  The four features considered are:  absolute value of the line-of-sight velocity $|v_{\mathrm{los}}|$, line-of-sight velocity normalized by velocity dispersion $|v_{\mathrm{los}}|/\sigma_v$, projected plane-of-sky position relative to halo radius $R_{\mathrm{sub}}/R_{200}$, and normalized effective angular momentum $|\mathrm{L}_{\mathrm{eff}}|=(\mathrm{R}_{\mathrm{sub}}/R_{200})|\mathrm{v}_{\mathrm{los}}|$.  

The first of the four ML feature sets (ML1, with $|v_{\mathrm{los}}|$ only) is chosen to mimic the data used in the $M(\sigma_v)$ power law (PL1), using only line-of-sight velocities to predict halo mass.  However, as was shown by \cite{Evrard:2008aa}, $\approx 10\%$ of halos are merger transients.  As an example, a line of sight merger of dissimilar-mass halos would exhibit notable skewness in the line-of-sight velocity PDF, with the sign of the skewness being indicative of whether the less massive halo was in the foreground or background of the more massive one.  In light of this, $|v_{\mathrm{los}}|$ is chosen as a feature over $v_{\mathrm{los}}$ for its virtue of making positive and negative skewness in the velocity PDF appear as identical systems.  ML1 mimics PL1, though going further by now predicting from the full LOS velocity distribution rather than a single summary statistic.  

ML2 also uses $|v_{\mathrm{los}}|$, but with the addition of the second feature, $|v_{\mathrm{los}}|/\sigma_v$.  The aim of explicitly normalizing the line-of-sight velocity distribution in this second feature is to highlight differences in $v_{\mathrm{los}}$ PDF shapes, mimicking the $\kappa$-corrected power law that was explored in PL2.

The third and fourth ML feature sets employ additional galaxy position information.  Knowing that $R_{200}$ correlates with halo mass, we choose $R_{\mathrm{sub}}/R_{200}$ as a way to utilize the relative distribution of galaxies without biasing the results by training on $R_{200}$.  ML3 uses $|v_{\mathrm{los}}|$ and $R_{\mathrm{sub}}/R_{200}$ as two separate features, while ML4 combines them into one feature, a normalized effective angular momentum $|\mathrm{L}_{\mathrm{eff}}|$.

\subsection{{Results}}

Figure \ref{fig:MLsummary} compares the predicted and actual masses for the clusters in our catalog, as well as an error comparison to PL1, the $M(\sigma_v)$ power law.  The bottom subpanel of each method comparison in the figure shows that the 68\% scatter is substantially decreased compared to PL1.  

The addition of $|v_{\mathrm{los}}|/\sigma_v$ in PL2 highlight{s} the difference in $v_{\mathrm{los}}$ PDF shape, much like the application of kurtosis did in PL2.  This may explain why, for most mass bins, the {mean $\epsilon$ is closer to zero, i.e.~the bias is smaller compared to} ML1 with the addition of this additional feature.  

However, comparing ML1 to ML3 with the addition of $R_{\mathrm{sub}}/R_{200}$ as a second feature, the {scatter in $\epsilon$ increases}.  To deduce the causes of this, we explore several variations to the ML3 method for comparison.  
We find that ML3 performs only {slightly} better than 
a feature set that uses $|v_{\mathrm{los}}|$ and a random number in place of the  $R_{\mathrm{sub}}/R_{200}$ value ($\epsilon \pm \Delta \epsilon = {-0.04^{+0.45}_{-0.31}}$).  From {this}, we conclude that the feature $R_{\mathrm{sub}}/R_{200}$, while containing information about the halo, tends to wash out the more important distribution of $|v_{\mathrm{los}}|$ in feature space.  
When the normalized $R_{\mathrm{sub}}/R_{200}$ is replaced with an unnormalized $R_{\mathrm{sub}}$, the result is similar to ML1 and ML2: $\epsilon \pm \Delta \epsilon = {-0.001^{+0.27}_{-0.22}}$.  
The quality of this fit is unsurprising because the maximum $R_{\mathrm{sub}}$ value is strongly correlated with $R_{200}$ and, therefore, with halo mass.  Despite the apparent failure of ML3, we remain optimistic that this feature could be of use when applied to a very large training catalog, when an appropriate simulation becomes available.

ML methods will tend to underpredict the most massive halo because the training set will include only halos less massive than this one outlier.  This effect is plainly evident for {ML2 and} ML3.
Because the mass predictions will tend to lie within the range of the training set masses, one should not overly interpret the 
prediction of the largest mass bin in Figure \ref{fig:MLsummary}, as it contains only the single most massive halo.

\section{Method Comparison \& Discussion}
\label{sec:discussion}

\begin{table*}[!htb]
\begin{center}
\caption{Method Comparison \label{table:methodcomp}}
\begin{tabular}{l l l l r r c c } 
\tableline
\tableline
\multicolumn{1}{l}{Case} & \multicolumn{1}{l}{Summary}  & \multicolumn{1}{l}{Color} & \multicolumn{1}{l}{Catalog} & \multicolumn{1}{c}{$\bar{\epsilon}$\,\tablenotemark{1}} & \multicolumn{1}{c}{$\epsilon \pm \Delta \epsilon$\,\tablenotemark{2}} & \multicolumn{1}{c}{$\Delta\epsilon$\,\tablenotemark{3}} & \multicolumn{1}{c}{$|\epsilon|-|\epsilon_{\textrm{PL1}}|$\,\tablenotemark{4}} \\
\tableline\\[1ex]

PL1 & $M(\sigma_v)$ Power Law& Red & Test &${0.128}$ & ${0.05^{+0.51}_{-0.36}}$ & ${0.871}$ & ---  \\[1.5ex]
 && & High-Mass Test & ${0.093}$ & ${0.02^{+0.44}_{-0.29}}$ & ${0.731}$ & --- \\[3.5ex]
 
PL2 & $M(\sigma_{\mathrm{v,\kappa\mbox{-}corrected}})$ Power Law & Blue & Test & ${0.079}$ &${0.04^{+0.41}_{-0.33}}$  &  ${0.735}$ & ${-0.06}$ \\[1.5ex]
&&& High-Mass Test & ${0.058}$ &${0.02^{+0.34}_{-0.27}}$ & ${0.612}$ &  ${-0.06}$ \\[3.5ex]

ML1 & SDM with $|v_{\mathrm{los}}|$  & Green & Test & ${-0.055}$ &${-0.07^{+0.25}_{-0.22}}$ & ${0.460}$ & ${-0.17}$  \\[1.5ex]
&&&High-Mass Test &${-0.042}$ &${-0.03^{+0.22}_{-0.18}}$ & ${0.402}$ & ${-0.14}$ \\[3.5ex]

ML2 &  SDM with $|v_{\mathrm{los}}|$ \& $|v_{\mathrm{los}}|/\sigma_v$ & Purple & Test & ${-0.038}$ & ${-0.05^{+0.25}_{-0.24}}$ & ${0.484}$ & ${-0.16}$ \\[1.5ex]
&&& High-Mass Test & ${-0.005}$ & ${-0.01^{+0.20}_{-0.19}}$ & ${0.386}$ & ${-0.16}$ \\[3.5ex]

ML3 & SDM with  $|v_{\mathrm{los}}|$ \& $R_{\mathrm{sub}}/R_{\mathrm{200}}$ & Orange & Test & ${0.005}$ &${-0.06^{+0.40}_{-0.28}}$ & ${0.679}$ &${-0.08}$ \\ [1.5ex]
&&& High-Mass Test & ${-0.017}$ &${-0.10^{+0.40}_{-0.25}}$ & ${0.651}$ &${-0.02}$ \\ [3.5ex]

ML4 & SDM with $|L_{\mathrm{eff}}|$ & Brown & Test & ${-0.066}$ &${-0.08^{+0.25}_{-0.22}}$ & ${0.468}$ & ${-0.16}$  \\[1.5ex]
&&& High-Mass Test & ${-0.063}$ &${-0.07^{+0.21}_{-0.20}}$ & ${0.410}$ & ${-0.13}$ \\[1.5ex]
\tableline
\end{tabular}

\tablenotetext{1}{Mean fractional mass error.}
\tablenotetext{2}{Median fractional mass error $\pm$ 68\% scatter.}
\tablenotetext{3}{Width of $\epsilon$ 68\% scatter.}
\tablenotetext{4}{Comparison of model to PL1; see equation \ref{eq:deltaepsilon}.}

\end{center}
\end{table*}

In this section, we will compare the six cluster mass prediction methods, using two different measures of comparison:  $\epsilon$ averaged across all clusters and $\epsilon$ as a function of mass.  

\begin{figure*}[!t]
\begin{center}
\begin{tabular}{c c}
     
     \footnotesize{Test Catalog} & \footnotesize{High-Mass Test Catalog}\\
     \includegraphics[width=0.5\textwidth]{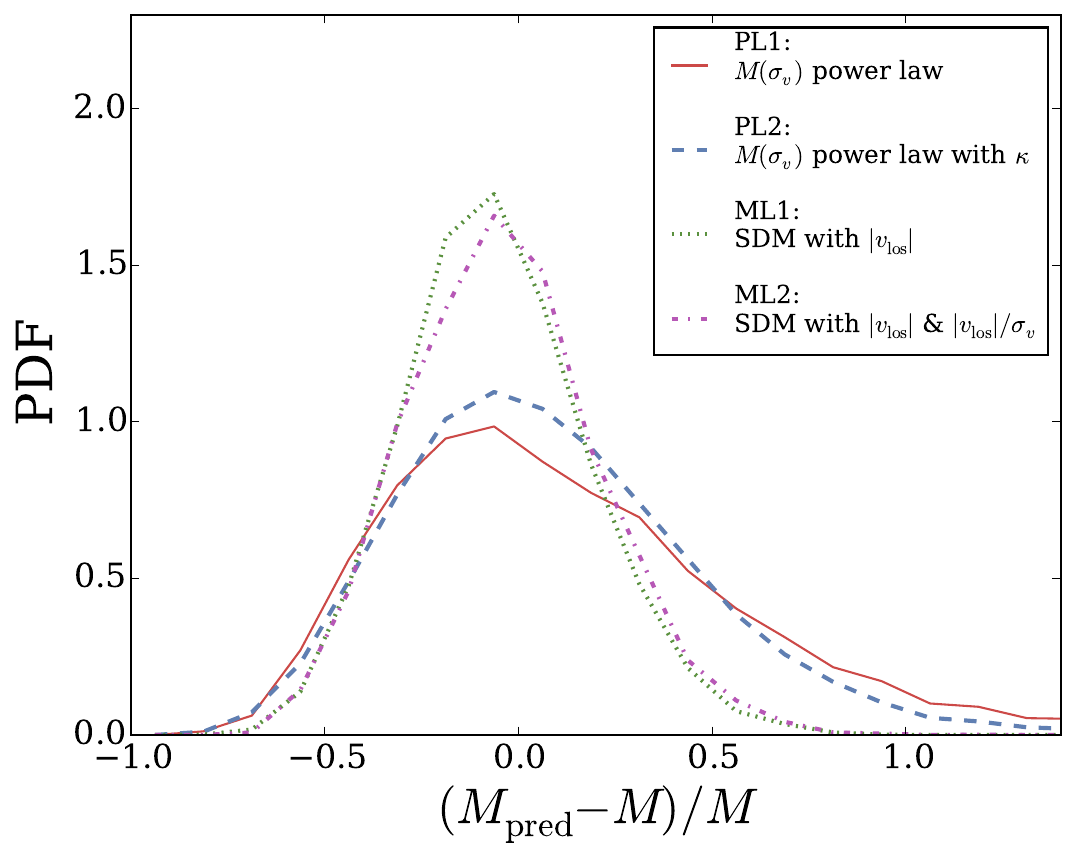} & \includegraphics[width=.5\textwidth]{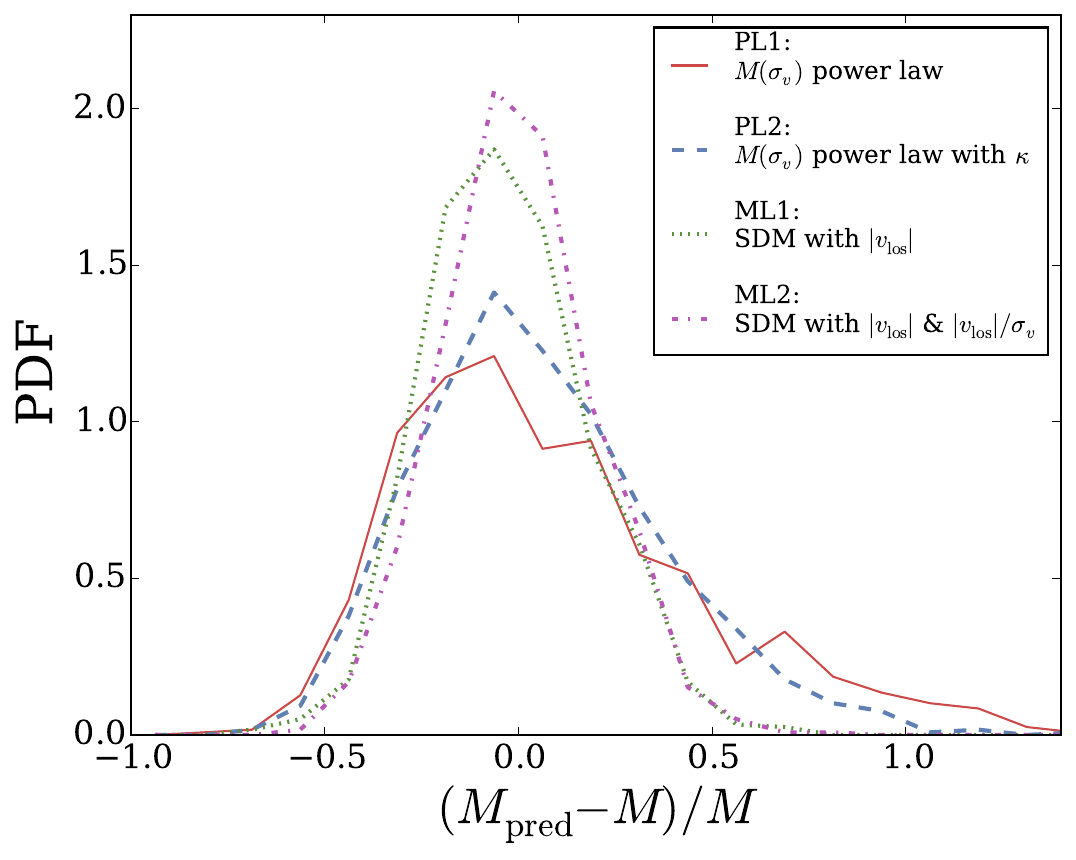} \\
         
  \end{tabular}
      \caption{PDF of the fractional mass error, $\epsilon=(M_{\textrm{pred}}-M)/M$, for four methods:  $M(\sigma_v)$ power law ({PL1, red solid}), $\kappa$-corrected power law ({PL2, blue dashed}), machine learning with $|v_{\mathrm{los}}|$ ({ML1, green dotted}), and machine learning with $|v_{\mathrm{los}}|$ and $|v_{\mathrm{los}}|/\sigma_v$ ({ML2, purple dash-dot}).  {In addition to the full Test Catalog population (left panel), the High-Mass Test Catalog ($M>7\times10^{14}\mathrm{M_{\odot}}h^{-1}$, right panel) is also shown.}
  PL2 shows a moderate improvement over PL1, while both ML methods shown outperform both power law methods, showing significantly smaller fractional mass errors. }
      \label{fig:fracErr}
      \end{center}
\end{figure*}

In Figure \ref{fig:fracErr}, which is a PDF of fractional mass errors, improvements in mass predictions are evident as we use more information from the LOS velocity PDF.  The addition of $\kappa$ as a predictor for residual $\delta$ in PL2 decreases the number of extreme overpredicted cluster masses and moderately improves mass prediction.  But the machine learning methods ML1 and ML2 significantly improve the accuracy of mass predictions.  With machine learning, the $\epsilon \gtrsim 0.6$ and $\epsilon \lesssim -0.6$ predictions are all but eliminated.  Machine learning clearly dominates by this measure:  averaged across all halos in the catalog, ML's cluster mass predictions are significantly improved over traditional power law predictions.

Table \ref{table:methodcomp} summarizes the mean fractional error ($\bar{\epsilon}$), median fractional error with 68\% scatter($\epsilon \pm \Delta \epsilon$), and width of the distributions ($\Delta\epsilon$) {for both the Test and High-Mass Test Catalogs.}  In this table, the mean and median $\epsilon$ quantify the bias: whereas the power law methods err on the side of overpredicting cluster masses, the machine learning methods err on the side of underprediction.  The $\Delta \epsilon$ value is a measure of the widths of the curves in Figure \ref{fig:fracErr}.  While PL1 {for the Test Catalog} has $\Delta\epsilon = 0.8{7}$, {PL2}
 decreases the width of the errors to 0.7{4}, a 1{6}\% improvement over the $M(\sigma_v)$ power law method.   Machine learning methods improve $\Delta\epsilon$ further:  ML3 ($\Delta\epsilon = 0.{68}$) improves {22}\% over the PL methods, while ML1 ($\Delta \epsilon=0.4{6}$), ML2 ($\Delta \epsilon=0.4{8}$), and ML4 ($\Delta \epsilon=0.4{7}$) have an even narrower distribution of $\epsilon$ values.  ML1, ML2, and ML4 have {47}\%, {44}\% and {46}\% improvements, respectively.  {Defining a fractional $\log$ mass error as ${\epsilon_{\log} = [\log(M_{\mathrm{pred}})-\log(M)]/\log(M)}$ in lieu of the fractional mass error $\epsilon$ produces similar comparitive results.}

\begin{figure*}[!t]
  \centering
  \includegraphics[width=\textwidth]{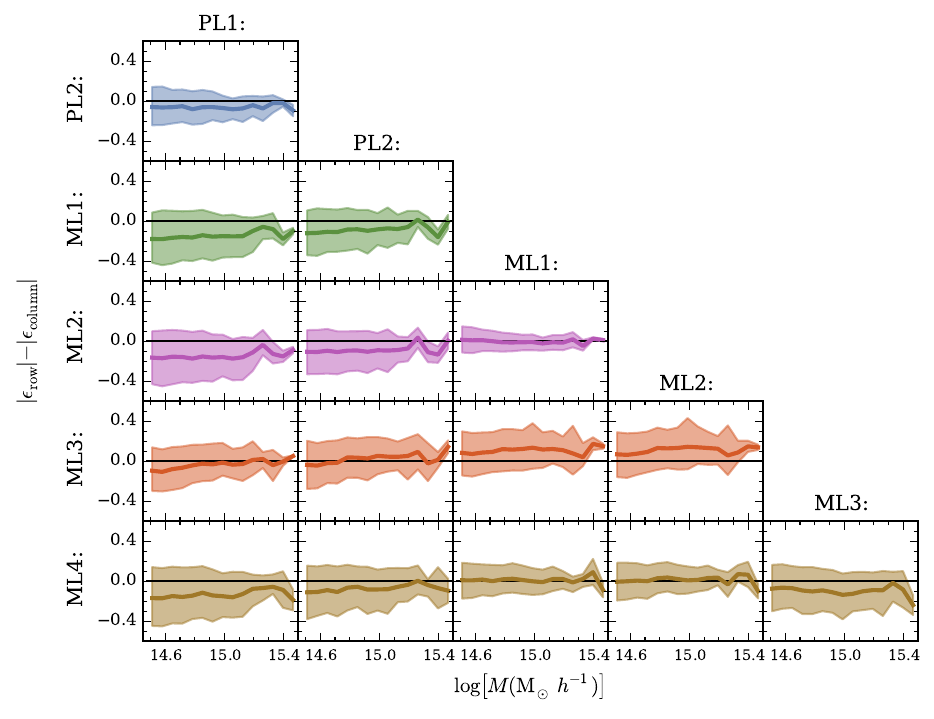}
  \caption{Method summary comparison: $|\epsilon_{\mathrm{row}}|-|\epsilon_{\mathrm{column}}|$ as a function of mass.  Values below 0 indicate that the row method is performing better than the column method for a given mass bin.  The left column summarizes a comparison of the five new methods to the $M(\sigma_v)$ power law:  PL2 (the $\kappa$-corrected power law) improves mass predictions in all mass bins.  While ML3, which includes $R_{\mathrm{sub}}/R_{\mathrm{200}}$, performs poorly compared to the other methods over much of the mass range considered, three machine learning methods---ML1, ML2, and ML4 (green, purple, and brown, respectively)---improve or maintain median accuracy of mass predictions {at all masses considered}.  }
  \label{fig:deltErr}
\end{figure*}

Figure \ref{fig:deltErr} is a direct comparison of the methods across mass bins. The comparison is quantified by
\begin{equation}
	|\epsilon_{\textrm{row}}|-|\epsilon_{\textrm{column}}|,
	\label{eq:deltaepsilon}
\end{equation}
where $\epsilon_{\textrm{column}}$ is the fractional mass error of the method indicated by the column label and $\epsilon_{\textrm{row}}$ is the fractional mass error of the row.  A $|\epsilon_{\textrm{row}}|-|\epsilon_{\textrm{column}}|$ value below zero is indicative of the row method predicting halo mass more accurately than the column method. 

The left column of Figure \ref{fig:deltErr} is a comparison to the $M(\sigma_v)$ power law, and the mean values of this comparison to PL1 are summarized in Table \ref{table:methodcomp}.  PL2 improves upon PL1 at all masses in the range considered, with an average $|\epsilon_{\textrm{PL2}}|-|\epsilon_{\textrm{PL1}}| = -0.0{6}$.  ML1 outperforms PL1 at all masses as well, but with a significantly smaller $|\epsilon_{\textrm{ML1}}|-|\epsilon_{\textrm{PL1}}| = -0.1{7}$.  ML2 and ML4 {also improve on PL1 at all masses, while ML3's improvement in mass predictions is the most pronounced at low masses.  At the highest masses, ML3 consistently underpredicts.  Recall, however,} that with machine learning methods, mass predictions typically lie within the range of the training set masses.  {Therefore,} we expect to see an underprediction for the most massive halo.  Table \ref{table:methodcomp} summarizes the $|\epsilon_{\textrm{ML}}|-|\epsilon_{\textrm{PL1}}|$ values for each method.

{The ML3 model underperforms at predicting most masses}.  This may be attributed to the inclusion of $R/R_{200}$, which  washes out the more important $|v_{\mathrm{los}}|$ feature, causing the significant underprediction of mass for {much of the mass range considered}.  ML3 presents a cautionary tale:  including additional distribution features with the SDM algorithm will not necessarily improve mass predictions, therefore, features should be chosen with care.

Because of its poor predictive power at {most masses}, ML3 is identified as a disfavored method.  Both ML1 and ML2, utilizing only constructs of the line-of-sight galaxy velocities, are our preferred machine learning methods.  Each has its own strength: ML1 outperforms ML2 {in the measure of ${\Delta\epsilon}$} for the Test Catalog, whereas ML2 {has a smaller ${\Delta\epsilon}$ for the High-Mass Test Catalog, outperforms in median ${\epsilon}$, and also minimizes the tendency to underpredict masses that is seen in the other machine learning methods.} 

{As tracers of the most massive dark matter halos, cluster counts as a function of mass and redshift are sensitive to cosmological parameters. Since they contain information about the growth of structure in the low-redshift universe, cluster richness and abundance have been used to constrain  constrain $\sigma_8$, $\Omega_m$, $\Omega_{\mathrm{DE}}$, and $w$ \citep[e.g.][]{2003A&A...398..867S, 2009ApJ...691.1307H, 2009ApJ...692.1060V, 2010ApJ...708..645R,2010MNRAS.406.1773M, 2011ARA&A..49..409A, 2014A&A...571A..20P}.  
For measurements of these types, cluster mass calculations remain a large source of systematic error. The improved errors, $\epsilon$, accessible using the machine learning technique discussed here could prove to be a powerful tool for addressing this problem. In principle, tighter constraints on scaling relations and the halo mass function could be made with the same number of observed clusters. Alternatively, fewer clusters would be needed in order to have the same amount of constraining power as current techniques. Ultimately, decreasing error in cluster mass measurements should result in more accurate values of cluster properties and cosmological parameters.}

\section{Conclusions}
\label{sec:conclusions}

We have explored dynamical mass measurements of a catalog of simulated galaxy clusters.  We present methods for cluster mass measurements that extract information from the line-of-sight velocity PDF, but improve upon the $M(\sigma_v)$ power law in accuracy.  All methods are trained on a catalog of simulated galaxy clusters with mass greater than $10^{14} \ \mathrm{M_{\odot}} h^{-1}$ and tested on those with mass greater than $3\times10^{14} \ \mathrm{M_{\odot}} h^{-1}$.  The halos used for building the cluster catalog are gleaned from a publicly-available halo catalog of the Multidark Simulation.

Two power law methods are considered:  PL1 employs a standard $M(\sigma_v)$ power law, while PL2 takes advantage of the relationship between residual ($\delta$) and LOS velocity PDF kurtosis ($\kappa$), using $\delta(\kappa)$ as a predictor for the amount by which the $M(\sigma_v)$ power law over- or underpredicts halo mass.  In addition to the power law method, we explore four machine learning methods, all of which employ Support Distribution Machines, a machine learning algorithm that learns from a distribution and predicts a mass.  For the ML methods, four different sets of distribution features are considered; all of these use only line-of-sight velocity, or LOS velocity coupled with normalized galaxy projected sky position, as raw data to construct the distribution features for training.  

Our main conclusions can be summarized as follows:
\begin{enumerate}

\item Applying the virial-theorem-motivated $M(\sigma_v)$ power law (method PL1) to our cluster catalog results in a fractional mass error width of $\Delta\epsilon = 0.8{71}$.

\item Kurtosis can be used to calculate a correction term for $\sigma_v$, resulting in a $\kappa$-corrected $M(\sigma_v)$ power law.  In some cases, it does so by identifying halos experiencing infalling matter by the negative $\kappa$ signature of the velocity PDF and appropriately lowering the prediction of these halos' masses.  
{This method (PL2),} decreases the width of fractional mass error to $\Delta \epsilon = 0.73{5}$, a decrease of 1{6}\% compared to the $M(\sigma_v)$ power law's predictions. 

\item Support Distribution Machines trained on one distribution feature, $|v_{\mathrm{los}}|$, (method ML1) results in $\Delta \epsilon = 0.4{60}$, a decrease of {47}\% compared to the $M(\sigma_v)$ power law's predictions.  {When SDM is} trained on two distribution features, $|v_{\mathrm{los}}|$ and $|v_{\mathrm{los}}/\sigma_v|$, (method ML2) results in $\Delta \epsilon = 0.4{84}$, a decrease of {44}\% compared to the $M(\sigma_v)$ power law's predictions.  Method ML2 is the preferred machine learning method for minimizing mean ${\epsilon}$.

\item Two additional SDM methods are tested, and their fractional error ($\epsilon$) values are summarized in Table \ref{table:methodcomp}.  Generally, these methods outperform both the $M(\sigma_v)$ power law method and kurtosis-corrected power law method (PL1 and PL2, respectively) {in terms of the width of fractional mass error $\Delta \epsilon$.  }

\end{enumerate}

In subsequent work, we will explore several remaining challenges in applying machine learning for dynamical mass measurements. {As discussed in \cite{2014MNRAS.441.1513O}, mass estimators perform best under the conditions for which they are calibrated; models that are calibrated to predict cluster mass in spite of the presence of interlopers actually tend to do worse when interlopers are entirely excluded from the sample and the true cluster membership is known.  Because the ultimate aim is to calibrate on simulations to predict masses of observed clusters, realistic} mock cluster catalogs that include known observational selection effects will have to be constructed and analyzed. Examples of major changes are: (1) including higher-redshift members to more accurately reflect the sample that will be probed by upcoming surveys, (2) assigning galaxy luminosities to subhalos and applying a galaxy luminosity cut rather than a subhalo mass cut, {(3) allowing for galaxies to be offset from the center of the halo or subhalo host, and (4)} incorporating observational selection criteria such as a fixed aperture with a LOS velocity cut that allows for interlopers.   Before this method can be applied to observation, there remains a need to train SDM on an observationally-aligned cluster catalog built from a large-volume, high-resolution simulation. Once such a simulation becomes available, Support Distribution Machines will be a powerful tool to predict cluster masses.

\acknowledgments{We thank {our referee\textbf{ Ramin Skibba for his constructive comments and review of this manuscript.  We also thank}} Alex Geringer-Sameth, {Shirley Ho,} Paul La Plante, Rachel Mandelbaum{, and Ying Zu} for their valuable feedback.
This work is supported in part by DOE DE-SC0011114 grant. 
The CosmoSim database used in this paper is a service by the Leibniz-Institute for Astrophysics Potsdam (AIP).
The MultiDark database was developed in cooperation with the Spanish MultiDark Consolider Project CSD2009-00064.
The Bolshoi and MultiDark simulations have been performed within the Bolshoi project of the University of California High-Performance AstroComputing Center (UC-HiPACC) and were run at the NASA Ames Research Center. The MultiDark-Planck (MDPL) and the BigMD simulation suite have been performed in the Supermuc supercomputer at LRZ using time granted by PRACE.}


\end{document}